\documentclass{elsart}
\usepackage{ifpdf}
\usepackage{graphicx,amssymb,lineno,epsfig,subfig}
\usepackage{subfig}
\usepackage{indentfirst}
\ifpdf
\usepackage[%
  pdftitle={Instructions for use of the document class
    elsart},%
  pdfauthor={Simon Pepping},%
  pdfsubject={The preprint document class elsart},%
  pdfkeywords={instructions for use, elsart, document class},%
  pdfstartview=FitH,%
  bookmarks=true,%
  bookmarksopen=true,%
  breaklinks=true,%
  colorlinks=true,%
  linkcolor=blue,anchorcolor=blue,%
  citecolor=blue,filecolor=blue,%
  menucolor=blue,pagecolor=blue,%
  urlcolor=blue]{hyperref}
\else
\usepackage[%
  breaklinks=true,%
  colorlinks=true,%
  linkcolor=blue,anchorcolor=blue,%
  citecolor=blue,filecolor=blue,%
  menucolor=blue,pagecolor=blue,%
  urlcolor=blue]{hyperref}
\fi

\renewcommand{\theequation}{\arabic{section}.\arabic{equation}}
\newtheorem{theorem}{Theorem}[section]

\newtheorem{example}{Example}[section]

\def\@subjclass{}
\makeatletter
\def\elsartstyle{%
    \def\normalsize{\@setfontsize\normalsize\@xiipt{14.5}}
    \def\small{\@setfontsize\small\@xipt{13.6}}
    \let\footnotesize=\small
    \def\large{\@setfontsize\large\@xivpt{18}}
    \def\Large{\@setfontsize\Large\@xviipt{22}}
    \skip\@mpfootins = 18\p@ \@plus 2\p@
    \normalsize
} \@ifundefined{square}{}{\let\Box\square} \makeatother
\usepackage{indentfirst}
\usepackage{setspace}
\begin{document}
\begin{frontmatter}
\title{Soliton solution of the osmosis $K(2, 2)$ equation}
\author{Jiangbo Zhou\corauthref{cor}},
\corauth[cor]{Corresponding author. Tel.: +86-511-88969336; Fax:
+86-511-88969336.} \ead{zhoujiangbo@yahoo.cn}
\author{Lixin Tian}
\address{Nonlinear Scientific Research Center, Faculty of Science, Jiangsu
University, Zhenjiang, Jiangsu 212013, China}
\begin{abstract} In this Letter, by using the bifurcation method of planar dynamical
systems, we obtain the analytic expressions of soliton solution of
the osmosis $K(2, 2)$ equation: $u_t+(u^2)_x-(u^2)_{xxx}=0$.

\end{abstract}


\end{frontmatter}
\section{Introduction}
\label{}
 \setcounter {equation}{0}

Since the theory of solitons has very wide applications in fluid
dynamics, nonlinear optics, biochemistry, microbiology, geophysics
and many other fields, the study of soliton solutions has become one
of the important issues of nonlinear evolution equations
\cite{1}-\cite{8}.

 Recently, Xu and Tian \cite{9} investigated the osmosis $K(2,2)$
equation
\begin{equation}
 \label {eq1.1}  u_t +(u^2)_x - (u^2)_{xxx}=0,
\end{equation}
where the positive convection term $(u^2)_x$ means the convection
moves along the motion direction, and the negative dispersive term
$(u^2)_{xxx}$ denotes the contracting dispersion. They obtained the
peaked solitary wave solution and the periodic cusp wave solution of
Eq.(\ref{eq1.1}). Unfortunately, the results are not complete. In
the present Letter, we shall continue their work and obtain the
smooth soliton solutions of Eq.(\ref{eq1.1}), so that we can
supplement the results of \cite{9}.

The remainder of this Letter is organized as follows. In Section 2,
we state the main results which are analytic expressions of the
smooth soliton solutions of Eq.(\ref{eq1.1}). In Section 3, we give
the proof of the main results.

\section{Main results}

 \setcounter {equation}{0}
 We state our main result as follows.
\begin{theorem} For given constants $c\neq 0$ and  $-\frac{c^2}{4}<g<-\frac{2c^2}{9}$, let $ \xi = x - ct$,

 (1) If $c<0$, then Eq.(\ref{eq1.1}) has a soliton solution of the form:
\begin{equation}
\label{eq2.1} \beta_1(\varphi _1^-)= \beta_1(\varphi) \exp({ -
\frac{1}{2}\xi }),
\end{equation}
where
\begin{equation}
\label{eq2.2} \beta_1(\varphi)=\frac{(2\sqrt {\varphi^2 + l_1
\varphi
 + l_2 } + 2\varphi  + l_1 )(\varphi  - \varphi _0^ - )^{\alpha_1
 }}{(2\sqrt {a_1 } \sqrt {\varphi^2 + l_1 \varphi + l_2 } +
b_1 \varphi  + l_3 )^{\alpha _1 }},
\end{equation}
\begin{equation}
\label{eq2.3}  \varphi _0^- =\frac{1}{2} ( c  - \sqrt {c^2 +4g}),
\end{equation}

\begin{equation}
\label{eq2.4}  \varphi _1^- =\frac{1}{6} ( c +3\sqrt {c^2 +4g} -
2\sqrt {c^2 +3c\sqrt {c^2 +4g}}),
\end{equation}
\begin{equation}
\label{eq2.5} l_1 = -\frac{1}{3}(c+3\sqrt {c^2 +4g}),
\end{equation}
\begin{equation}
\label{eq2.6} l_2 = \frac{1}{6}(c^2+6g-c\sqrt {c^2 +4g}),
\end{equation}
\begin{equation}
\label{eq2.7} l_3 =  \frac{2}{3}(c^2+6g-c\sqrt {c^2 +4g}),
\end{equation}
\begin{equation}
\label{eq2.8} a_1 = c^2+4g-c\sqrt {c^2 +4g},
\end{equation}
\begin{equation}
\label{eq2.9} b_1 =\frac{1}{3}(2c-6\sqrt {c^2 +4g}),
\end{equation}
\begin{equation}
\label{eq2.10} \alpha _1 = \frac{c-\sqrt {c^2 +4g}}{2\sqrt {c^2
+4g-c\sqrt {c^2 +4g}}}.
\end{equation}
\noindent (2) If $c>0$, then Eq.(\ref{eq1.1}) has a soliton solution
of the form:
\begin{equation}
\label{eq2.11} \beta_2(\varphi )= \beta_2(\varphi_1^+) \exp({
\frac{1}{2}\xi }),
\end{equation}
where
\begin{equation}
\label{eq2.12} \beta_2(\varphi)=\frac{(2\sqrt {\varphi^2 + m_1
\varphi
 + m_2 } + 2\varphi  + m_1 )( \varphi _0^ + -\varphi )^{\alpha_2
 }}{(2\sqrt {a_2 } \sqrt {\varphi^2 + m_1 \varphi + m_2 } +
b_2 \varphi  + m_3 )^{\alpha _2 }},
\end{equation}
\begin{equation}
\label{eq2.13}  \varphi _0^+ =\frac{1}{2} ( c  + \sqrt {c^2 +4g}),
\end{equation}

\begin{equation}
\label{eq2.14}  \varphi _1^+ =\frac{1}{6} ( c +3\sqrt {c^2 +4g} +
2\sqrt {c^2 +3c\sqrt {c^2 +4g}}),
\end{equation}

\begin{equation}
\label{eq2.15} m_1 = -\frac{1}{3}(c-3\sqrt {c^2 +4g}),
\end{equation}
\begin{equation}
\label{eq2.16} m_2 = \frac{1}{6}(c^2+6g+c\sqrt {c^2 +4g}),
\end{equation}
\begin{equation}
\label{eq2.17} m_3 =  \frac{2}{3}(c^2+6g+c\sqrt {c^2 +4g}),
\end{equation}
\begin{equation}
\label{eq2.18} a_2 =c^2+4g+c\sqrt {c^2 +4g},
\end{equation}
\begin{equation}
\label{eq2.19} b_2 =\frac{1}{3}(2c+6\sqrt {c^2 +4g}),
\end{equation}
\begin{equation}
\label{eq2.20} \alpha _2 = \frac{c+\sqrt {c^2 +4g}}{2\sqrt {c^2
+4g+c\sqrt {c^2 +4g}}}.
\end{equation}
\end{theorem}

We shall give the proof of this theorem in Section 3. Now we take a
set of data and employ Maple to display the graphs of the soliton
solutions on $u -\xi $ plane.

\begin{example}

 Taking $c = -2$ and $g =-0.9999$ (corresponding to (1) of
Theorem 2.1), it follows that $\varphi _0^ - = -1.01$, $\varphi_1^-
= -0.979924$, $l_1 = 0.646667$ and $l_2 = -0.326567$, $l_3 =
-1.30627$, $a_1 = 0.0404$, $b_1 = -1.37333$, and $\alpha _1 =
-5.02494$. We present the graph of the soliton solution in Fig.1
(a).
\end{example}

\begin{example}
 Taking $c = 2$ and $g =-0.9999$ (corresponding to (2) of
Theorem 2.1), it follows that $\varphi _0^+ =1.01$, $\varphi_1^+ =
0.979924$, $m_1 = -0.646667$ and $m_2 = -0.326567$, $m_3 =
-1.30627$, $a_2 =0.0404$, $b_2 = 1.37333$, $\alpha _2 = 5.02494$.
The graph of the soliton solution is presented in Fig.1(b).
\end{example}

\begin{figure}[h]
\centering \subfloat[]{\label{fig:1}
\includegraphics[height=1.3in,width=2.2in]{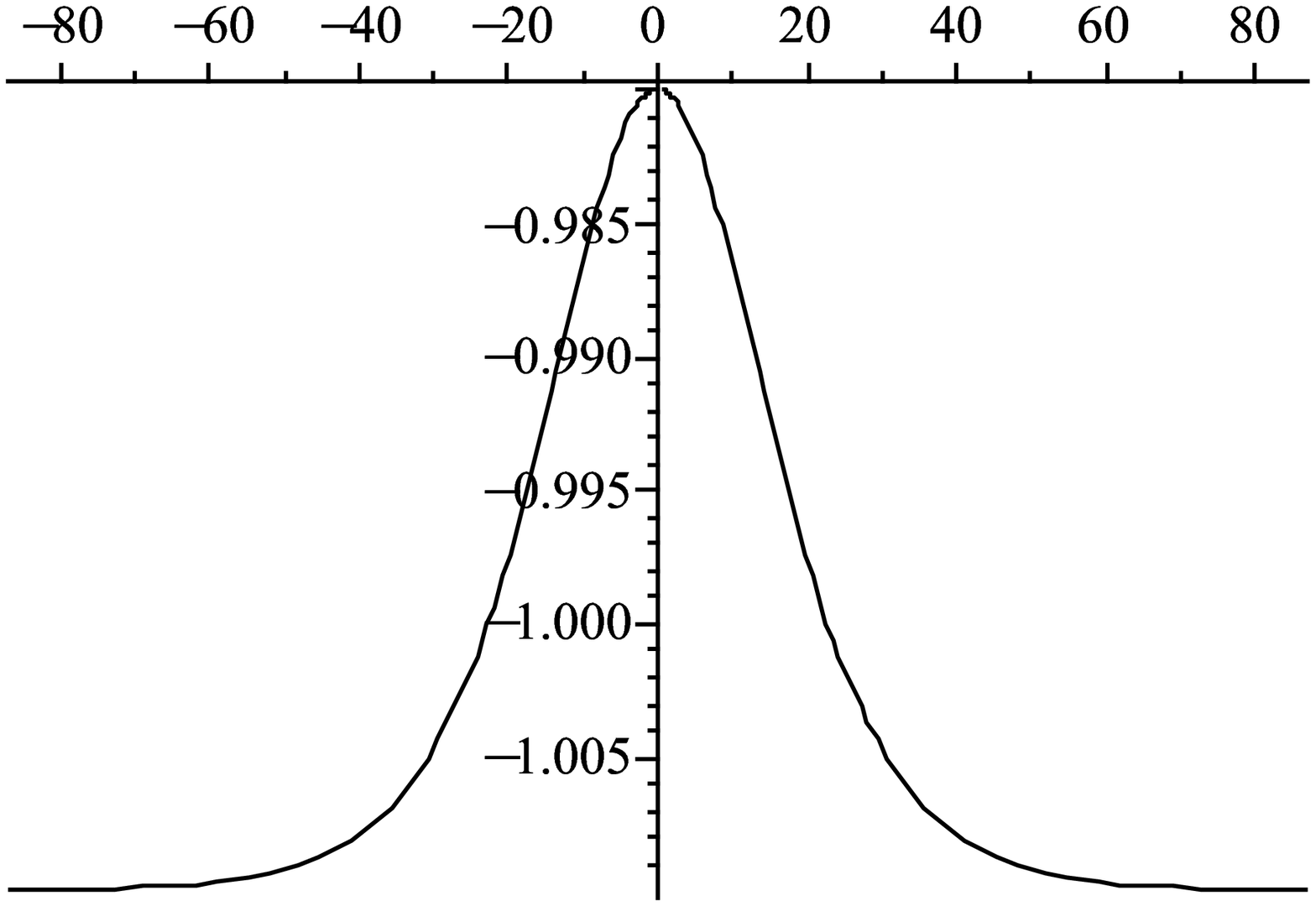}}\hspace{0.1\textwidth}
\subfloat[ ]{ \label{fig:2}
\includegraphics[height=1.3in,width=2.2in]{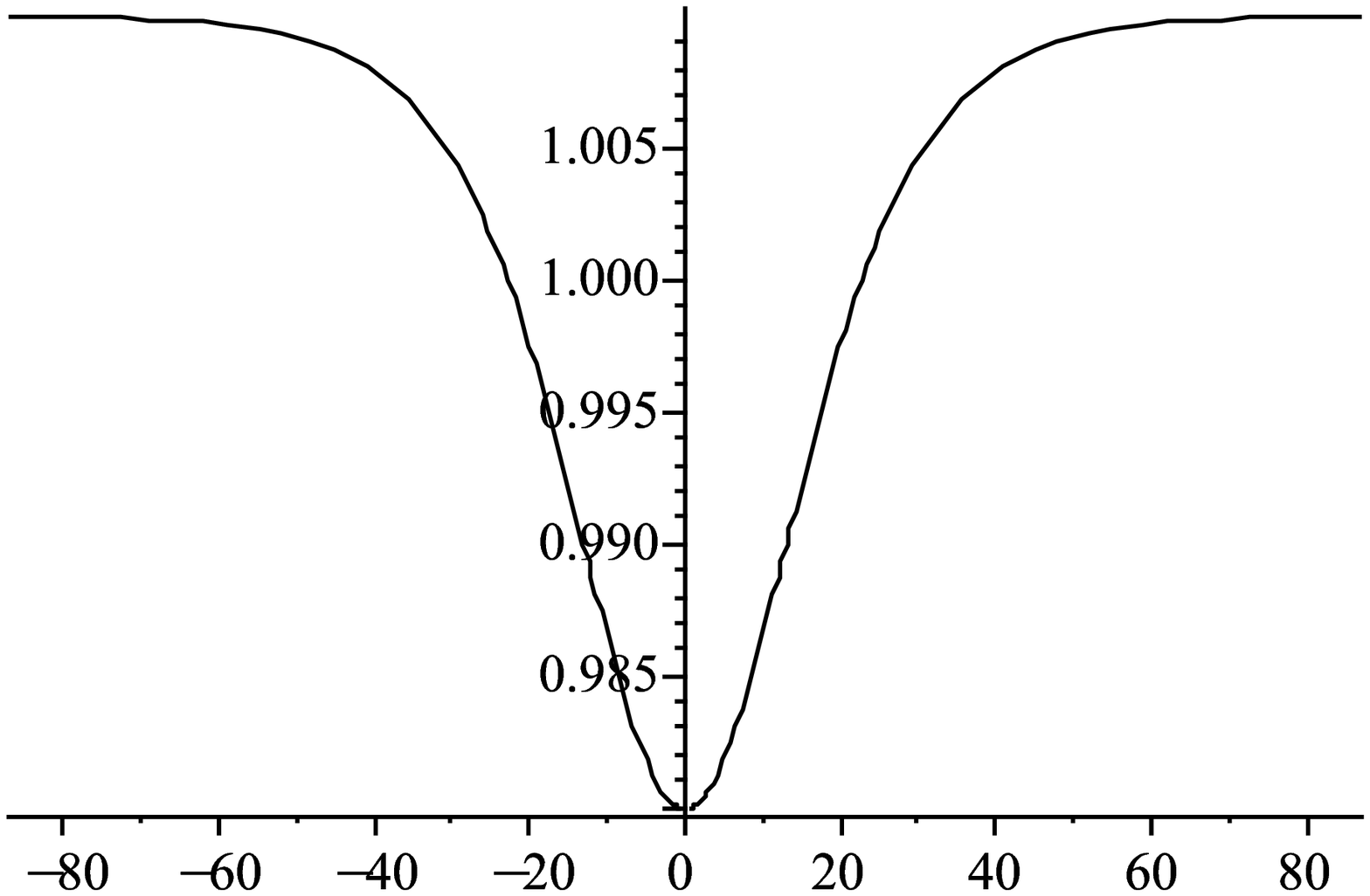}}

\caption{The graphs of the soliton solutions. (a) $c=-2$,
$g=-0.9999$; (b) $c=2$, $g=-0.9999$.}
\end{figure}

\section{ Proof of main results}
\setcounter {equation}{0}
 Eq.(\ref{eq1.1}) also takes the form
\begin{equation}
 \label {eq3.1}  u_t +2uu_x - 6u_xu_{xx}-2uu_{xxx}=0,
\end{equation}

Let $u = \varphi (\xi )$   with $\xi = x - ct (c\neq0)$  be the
solution of Eq.(\ref{eq3.1}), then it follows that
 \begin{equation}
\label{eq3.2}
 - c\varphi' + 2\varphi \varphi '-6\varphi \varphi '' -2\varphi \varphi
 '''=0.
\end{equation}

Integrating (\ref{eq3.2}) once we have
\begin{equation}
\label{eq3.3} - c\varphi + (\varphi)^2-2( \varphi ')^2-2\varphi
\varphi '' =g,
\end{equation}
\noindent where $g$ is the integral constant.

Let $y = \varphi '$, then we get the following planar dynamical
system:
\begin{equation}
\label{eq3.4} \left\{ {\begin{array}{l}
 \frac{d\varphi }{d\xi } = y \\
 \frac{dy}{d\xi } = \frac{\varphi ^2-c \varphi-g -2y^2}{2\varphi}\\
 \end{array}} \right.
\end{equation}
\noindent with a first integral
\begin{equation}
\label{eq3.5} H(\varphi, y)=\varphi^2(y^2 -
\frac{1}{4}\varphi^2+\frac{c}{3}\varphi+ \frac{1}{2} g )=h,
\end{equation}
\noindent where $h$ is a constant.

Note that (\ref{eq3.4}) has a singular line $\varphi = 0$, to avoid
the line temporarily we make transformation $d\xi = 2\varphi d\zeta
$. Under this transformation, Eq.(\ref{eq3.4}) becomes
\begin{equation}
\label{eq3.6} \left\{ {\begin{array}{l}
 \frac{d\varphi }{d\zeta } =  2\varphi y \\
 \frac{dy}{d\zeta } = \varphi ^2-c \varphi-g -2y^2
\\
 \end{array}} \right.
\end{equation}
Eq.(\ref{eq3.4}) and Eq.(\ref{eq3.6}) have the same first integral
as (\ref{eq3.5}). Consequently, system (\ref{eq3.4}) has the same
topological phase portraits as system (\ref{eq3.6}) except for the
straight line $\varphi = 0$. Obviously, $\varphi = 0$ is an
invariant straight-line solution of system (\ref{eq3.6}).

For a fixed $h$, (\ref{eq3.5}) determines a set of invariant curves
of (\ref{eq3.6}). As $h$ is varied, (\ref{eq3.5}) determines
different families of orbits of (\ref{eq3.6}) having different
dynamical behaviors. Let $M(\varphi _e ,y_e )$ be the coefficient
matrix of the linearized system of (\ref{eq3.6}) at the equilibrium
point $(\varphi _e ,y_e )$, then

\[
M(\varphi _e ,y_e ) = \left( {{\begin{array}{*{20}c}
{\indent  y_e } &&& {\indent 2\varphi _e}   \\
 {2\varphi _e - c }   &&& \indent{- 4y_e}   \\
\end{array} }} \right)
\]
\noindent and at this equilibrium point, we have
\[
 J(\varphi _e ,y_e ) = \det M(\varphi _e ,y_e ) = - 4y_e^2 -
4\varphi _e (\varphi _e -\frac {c}{2} ),
\]
\[
  p(\varphi _e ,y_e ) =
\mathrm{trace}(M(\varphi _e ,y_e )) = - 3y_e. \nonumber
\]
By the theory of planar dynamical system (see \cite {10}), for an
equilibrium point of a planar dynamical system, if $J < 0$, then
this equilibrium point is a saddle point; it is a center point if $J
> 0$ and $p = 0$; if $J = 0$ and the Poincar\'{e} index of the
equilibrium point is 0, then it is a cusp.

In \cite{9}, Xu and Tian reported that when
$-\frac{c^2}{4}<g<-\frac{2c^2}{9}$,  system (\ref{eq3.4})  has two
equilibrium points $(\varphi _0^ - ,0)$ and $(\varphi _0^ + ,0)$.
(i) If $c<0$, then $(\varphi _0^ - ,0)$ is a saddle point while
$(\varphi _0^ + ,0)$ is a center point. There is inequality $\varphi
_0^ - <\frac{c}{2}<\varphi _0^ +<0$; (ii) If $c>0$ , then $(\varphi
_0^ - ,0)$ is a center point while $(\varphi _0^ + ,0)$ is a saddle
point. There is inequality $0<\varphi _0^ - <\frac{c}{2}<\varphi _0^
+$.

In the parameter region: $-\frac{c^2}{4}<g<-\frac{2c^2}{9}$, we show
the phase portraits of system (\ref{eq3.4}) in Fig.2.

\begin{figure}[h]
\centering \subfloat[]{\label{fig:1}
\includegraphics[height=1.8in,width=2.2in]{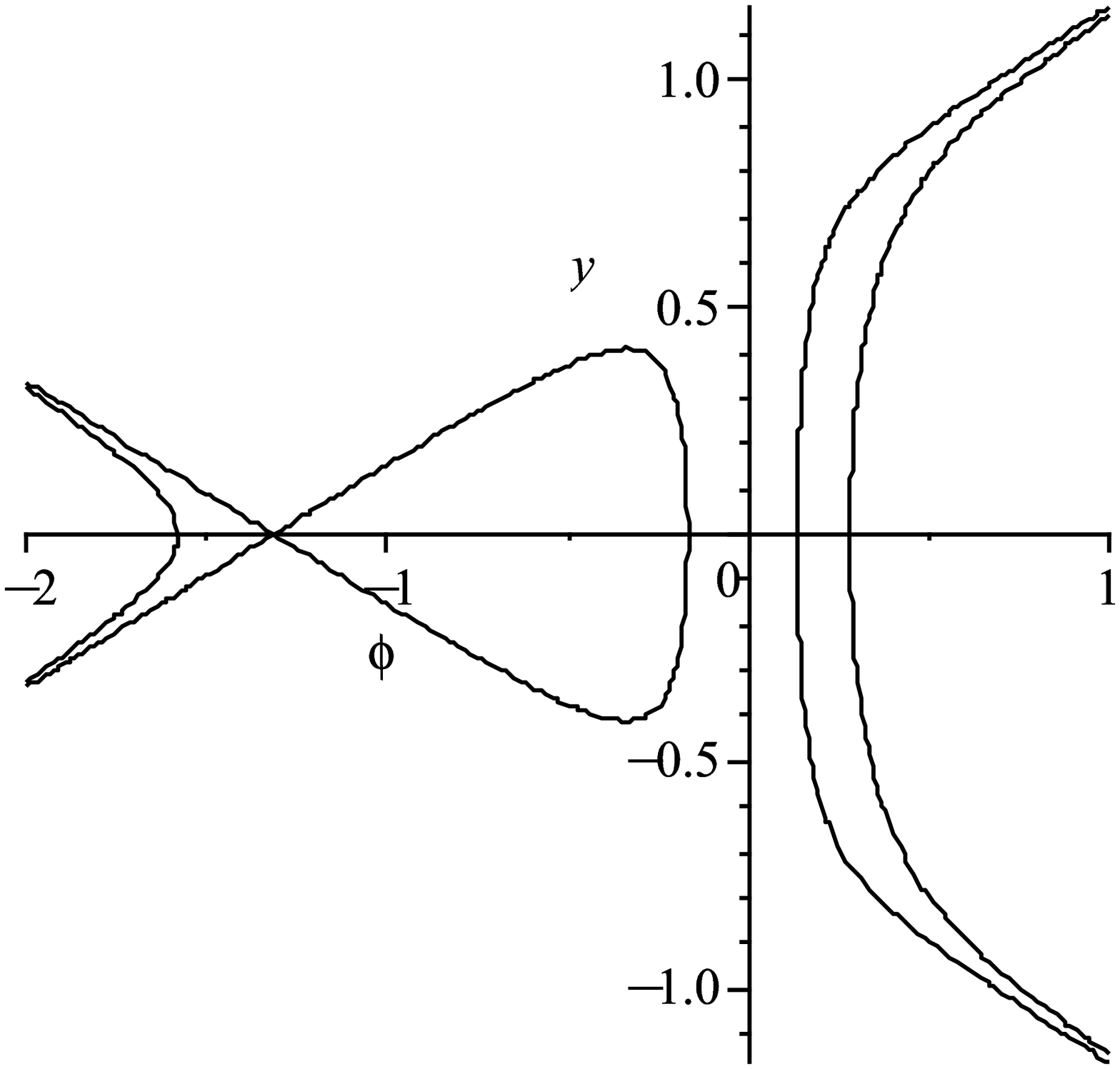}}\hspace{0.1\textwidth}
\subfloat[ ]{ \label{fig:2}
\includegraphics[height=1.8in,width=2.2in]{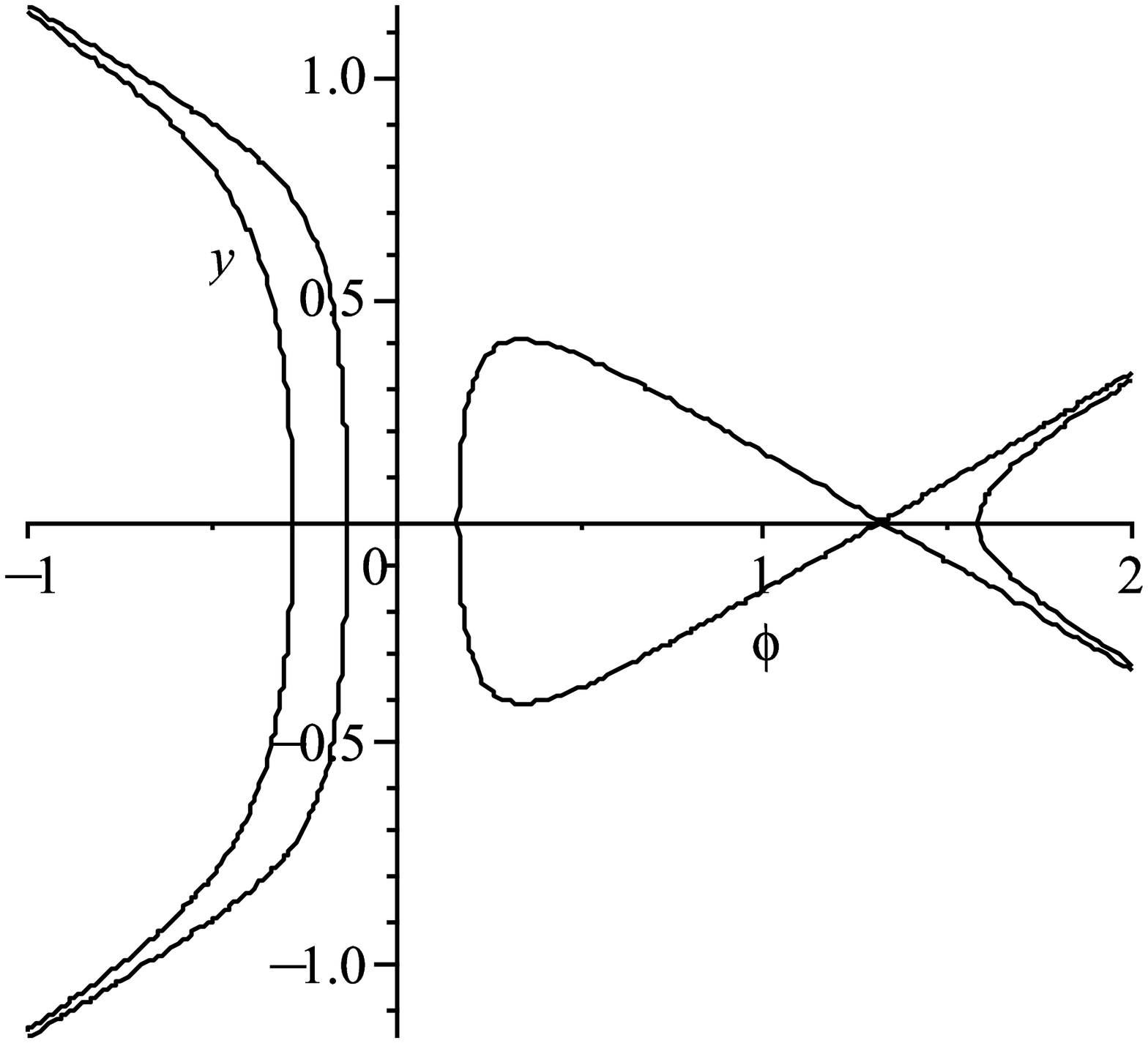}}

\caption{The hase portraits of system (\ref{eq3.4}) when
$-\frac{c^2}{4}<g<-\frac{2c^2}{9}$. (a) $c<0$; (b) $c>0$.}
\end{figure}

Usually, a soliton solution of Eq.(\ref{eq1.1}) corresponds to a
homoclinic orbit of system (\ref{eq3.4}). In Fig.2(a), the
homoclinic orbit of system (\ref{eq3.4}) can be expressed as
\begin{equation}
\label{eq3.7}  y = \pm  \frac{(\varphi -\varphi _0^ - )\sqrt
{\varphi ^2 + l_1 \varphi + l_2 } }{2\varphi} \quad for \quad
\varphi _0^ -\leq \varphi \leq \varphi _1^-,
\end{equation}
where $\varphi_0^-$, $\varphi_1^-$, $l_1$ and $l_2$ are in
(\ref{eq2.3}), (\ref{eq2.4}), (\ref{eq2.5}) and (\ref{eq2.6}),
respectively. Substituting Eq.(\ref{eq3.7}) into the first equation
of system (\ref{eq3.4}) and integrating along the homoclinic orbits,
we have the expression of the solion solution as in (\ref{eq2.1}).

In Fig.2(b), the homoclinic orbit of system (\ref{eq3.4}) can be
expressed as
\begin{equation}y=\pm \frac{(\varphi -
\varphi _0^ + )\sqrt {\varphi ^2 + m_1 \varphi + m_2 } }{2\varphi}
\quad for \quad \varphi _1^ +\leq \varphi \leq \varphi _0^+,
\end{equation}
where $\varphi_0^+$,  $\varphi_1^+$, $m_1$ and $m_2$ are in
(\ref{eq2.13}), (\ref{eq2.14}), (\ref{eq2.15}) and (\ref{eq2.16}),
respectively. Substituting Eq.(\ref{eq3.7}) into the first equation
of system (\ref{eq3.4}) and integrating along the homoclinic orbits,
we have the expression of the soliton solution as in (\ref{eq2.11}).
The proof of Theorem 2.1 is completed.

\end{document}